\documentclass[10pt,letterpaper]{article}
\usepackage{opex3}
\usepackage{braket,amsmath,mathrsfs}

\newcommand*\op[1]{\ensuremath \hat{#1}}
\newcommand*\opd[1]{\op{#1}^\dagger}
\renewcommand*\H{\op{\mathscr{H}}}

\newcommand*\ii{\ensuremath i}
\newcommand*\twovec[2]{\begin{pmatrix} #1 \\ #2 \end{pmatrix}}
\newcommand*\tworvec[2]{\begin{pmatrix} #1, & #2 \end{pmatrix}}
\newcommand*\expect[1]{\ensuremath\langle#1\rangle}
\newcommand*{\var}[1]{\ensuremath \text{Var}\left(#1\right)}
\newcommand*{\cov}[2]{\ensuremath \text{CoVar}\left(#1,\,#2\right)}
\newcommand*\mat[1]{\ensuremath \mathbf{#1}}
\newcommand*\ord[2]{O\left(\tfrac{#1}{#2}\right)}

\usepackage{cite}
\usepackage[draft]{hyperref}

\begin{document}

\title{Relative intensity squeezing by four-wave mixing with loss:
an analytic model and experimental diagnostic}

\author{M.~Jasperse,$^{1,2,*}$ L.~D.~Turner,$^2$ and R.~E.~Scholten$^1$}

\address{$^1$ARC Centre of Excellence for Coherent X-Ray Science, University of Melbourne,\\
VIC 3010, Australia\\$^2$ School of Physics, Monash University,
VIC 3800, Australia}

\email{*martijn.jasperse@monash.edu}

\begin{abstract}
Four-wave mixing near resonance in an atomic vapor can produce relative intensity squeezed light suitable for precision measurements beyond the shot-noise limit. 
We develop an analytic distributed gain/loss model to describe the
competition of mixing and absorption through the non-linear
medium. Using a novel matrix calculus, we present closed-form
expressions for the degree of relative intensity squeezing
produced by this system. We use these theoretical results to
analyze experimentally measured squeezing from a $^{85}$Rb vapor
and demonstrate the analytic model's utility as an experimental
diagnostic.
\end{abstract}


\section{Introduction}
Relative intensity squeezing by four-wave mixing in an atomic
vapor is emerging as a promising technique for performing
high-precision measurements beyond the shot-noise limit. First
demonstrated by McCormick et al.~\cite{lett07}, the technique uses
atomic coherences to produce quantum correlated ``twin beams'',
enabling the shot-noise of one beam to be measured and subtracted
from the other to obtain a low-noise differential measurement; for
example of a weakly absorbing sample. This scheme was recently
shown to reduce the relative intensity noise by $9.2\pm0.5\,$dB
below the shot-noise limit~\cite{glorieux}, and noise reduction
has been observed in both the low Fourier frequency
\cite{lett_low} and multi-mode imaging
\cite{lett_qimaging,lett_science} domains.

Furthermore, as one of the twin beams is near-resonant with the
atoms, this squeezing technique has promising applications in
quantum information processing \cite{lett_slow,lett_qpi}. However,
absorption near resonance degrades the quantum correlations. Both
mixing gain and absorption losses occur simultaneously as the
beams propagate through the vapor, and are therefore competing
processes.

Earlier theoretical investigations of this system have applied
numerical methods~\cite{lett_low} and the Heisenberg-Langevin
formalism~\cite{glorieux2} to predict the resulting degree of
squeezing. The numerical model demonstrated excellent agreement
with experimental results, but it can be difficult to gain insight
into the competing processes from numerical calculations. The
Heisenberg-Langevin model provided a microscopic description of a
specific four-wave mixing configuration in a cold atomic gas,
which accurately predicted the resulting gain profiles. However,
calculation of the predicted squeezing required complex matrix
integrals and no comparison to experimentally measured squeezing
was presented.

In this work, we present a very general approach for determining
the squeezing produced by a four-wave mixing system, and develop a
matrix-based analysis method to include arbitrarily many injected
vacuum modes. Considering special cases, simple closed-form
expressions are easily obtained. Finally, we present
experimentally measured squeezing from four-wave mixing in a
rubidium-85 vapor, and demonstrate how the model can be used as a
diagnostic tool to determine the limiting technical factors.

\section{Relative intensity squeezing}

\begin{figure}[!hbt]\centering
    \includegraphics[width=\textwidth]{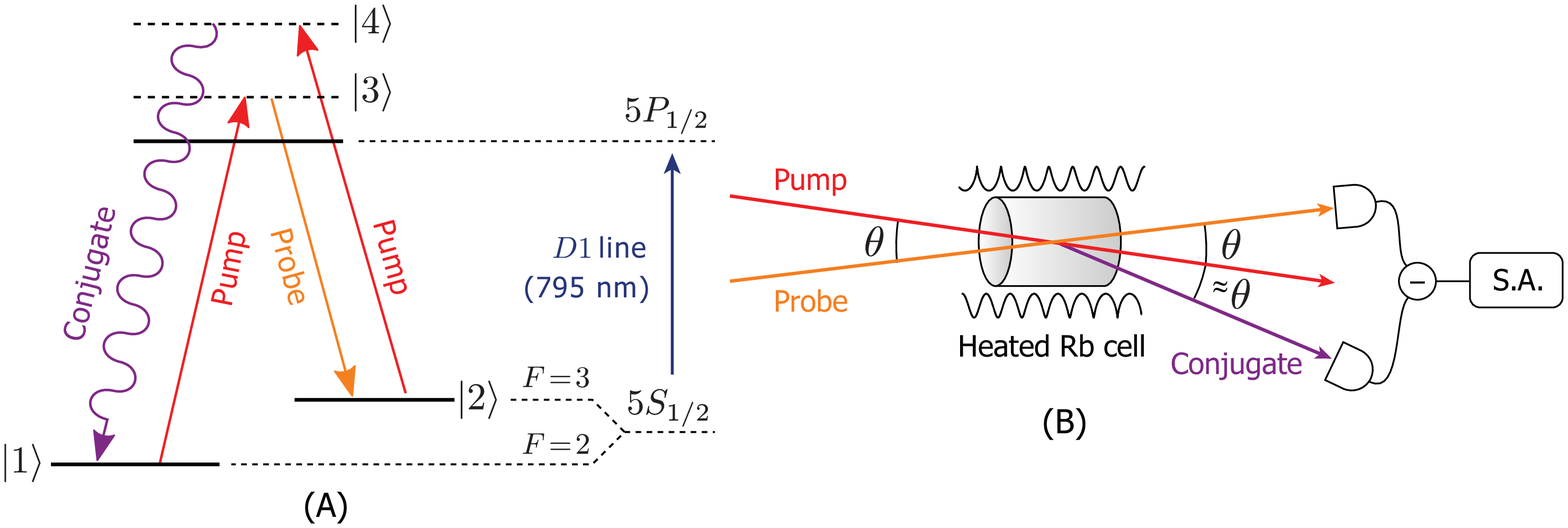}
\caption{(A) Four-wave mixing energy-level transitions and (B) and
experimental schematic.}
    \label{fig:fwm}
\end{figure}

The ``double-$\Lambda$'' four-wave mixing scheme introduced by
McCormick et al~\cite{lett07} uses a high-intensity ``pump'' beam
to drive a cycle of four off-resonant transitions in a hot
rubidium vapor, causing the emission of correlated ``probe'' and
``conjugate'' photons (Fig. \ref{fig:fwm}A). The probe transition
is stimulated by a seed laser incident at an angle $\theta$ to the
pump, resulting in the spontaneous emission of the conjugate on
the opposite side of the pump beam (Fig. \ref{fig:fwm}B). The beam
powers are measured individually and subtracted to obtain the
relative intensity noise as measured on a spectrum analyzer
(S.A.).

Labelling the Fock-space annihilation operators of the probe,
conjugate and pump by $\op{a}$, $\op{b}$ and $\op{c}$ respectively
and the interaction strength by $\xi$, the interaction picture
Hamiltonian is
\begin{equation*}
    \H_i = \ii\hbar (\xi \opd{b}\op{c}\opd{a}\op{c} - \xi^* \opd{c}\op{a}\opd{c}\op{b}) .
\end{equation*}
In the ``undepleted pump'' approximation, the intense pump beam
remains in its initial coherent state $\ket{\psi_c}$ and the
substitution $\op{c}\rightarrow\psi_c$ can be made:
\begin{equation*}
    \H_i = \ii\hbar (\xi \psi_c^2 \opd{b}\opd{a} - \xi^*(\psi_c^*)^2 \op{a}\op{b}) .
\end{equation*}
The time-evolution of this Hamiltonian over the interaction
time-scale $\tau$ is
\begin{equation}
    \op{S} \equiv \exp(-\ii \H_i \tau / \hbar) = \exp(s \opd{b}\opd{a} - s^*\op{a}\op{b}),      \qquad\mbox{where } s = \xi\psi_c^2\tau .
\end{equation}
This is the two-mode squeezing operator for modes $\op{a}$ and
$\op{b}$, where $s$ is the ``squeezing parameter''~\cite{knight}.
The four-wave mixing system therefore produces a two-mode squeezed
state, reducing amplitude difference noise at the expense of
increasing phase difference noise \cite{lett_science}.

The phase of $s$ results in a rotation of the (arbitrary)
measurement quadratures, so $s$ may be taken as real and positive.
The probe and conjugate modes $\op{a}$ and $\op{b}$ are then
transformed as
\begin{equation}
    \op{a} \rightarrow \opd{S}\op{a}\op{S} = \cosh(s) \op{a} + \sinh(s) \opd{b} \quad\text{ and }\quad
    \opd{b} \rightarrow \opd{S}\opd{b}\op{S} = \sinh(s) \op{a} + \cosh(s) \opd{b} .
    \label{eq:4wm}
\end{equation}
Defining the number operator of the incident probe beam as
$\op{N}_0\equiv\opd{a}_0\op{a}_0$ and making the bright beam
approximation $\expect{\op{N}_0}\gg1$, the number operators after
squeezing become
\begin{equation*}\begin{aligned}
    \expect{\op{N}_a} \equiv \expect{\opd{a}\op{a}} \simeq G\expect{\op{N}_0} \quad\text{ and }\quad
    \expect{\op{N}_b} \equiv \expect{\opd{b}\op{b}} \simeq (G-1)\expect{\op{N}_0} ,
\end{aligned}\end{equation*}
where $G\equiv\cosh^2 s$ is the increase in probe intensity,
termed the ``mixing gain''.

The relative intensity operator $\op{N}_a-\op{N}_b$ is unchanged
by $\op{S}$, so \mbox{$\var{\op{N}_a-\op{N}_b} = \var{\op{N}_0}$}.
Hence the beams have been amplified without increasing the
relative intensity noise; they are relative intensity squeezed.
The noise figure of the process (or ``degree of squeezing'') is
the ratio of the measured noise to the corresponding shot-noise
level for equal optical power. Assuming an initially shot-noise
limited probe, the noise figure is
\begin{equation}
    \mbox{NF} \equiv \frac{\var{\op{N}_a-\op{N}_b}}{\expect{\op{N}_a}+\expect{\op{N}_b}} = \frac{\expect{\op{N}_0}}{G\expect{\op{N}_0}+(G-1)\expect{\op{N}_0}} = \frac1{2G-1} .
\end{equation}
Therefore the measured noise power can be reduced arbitrarily
below the shot-noise limit in the limit of ideal detection.
However, optical losses are unavoidable and occur both within the
medium (e.g. absorption) and after it (e.g. imperfect detection).
These losses randomly eject photons from the probe and conjugate
beams, decorrelating their intensities and degrading the observed
degree of squeezing. We now construct models to quantify this
effect.

\section{Optical losses after squeezing}
We initially consider only losses that occur after mixing, such as
from imperfect optical transmission or detection efficiency. These
losses are modelled by a beamsplitter with an empty
port~\cite{bachor} whose output state is a combination of the
input and vacuum modes, contributing ``vacuum fluctuations'' to
the transmitted beam~\cite{caves80}. Denoting the vacuum modes
introduced by losses on the probe and conjugate by the
annihilation operators $\op{x}$ and $\op{y}$ respectively, the
standard beam-splitter input-output relations~\cite{loudon} give
\begin{equation}
    \op{a} \rightarrow \sqrt{\eta_a}\;\op{a} + \sqrt{1-\eta_a}\;\op{x}  \quad\text{ and }\quad
    \op{b}\rightarrow\sqrt{\eta_b}\;\op{b} + \sqrt{1-\eta_b}\;\op{y},
\label{eq:loss}
\end{equation}
where $\eta_a$ and $\eta_b$ are the fractions of the probe and
conjugate intensities transmitted. The relative intensity noise
can then be expressed in terms of the individual beam variances
and covariance to give
\begin{equation*}\begin{aligned}
    \var{\op{N}_a-\op{N}_b} &= \eta_a^2 \var{\op{N}_a} + \eta_a(1-\eta_a)\expect{\op{N}_a}  + \eta_b^2\var{\op{N}_b} + \eta_b(1-\eta_b)\expect{\op{N}_b} \\
        & \quad\qquad - 2\eta_a\eta_b\cov{\op{N}_a}{\op{N}_b} .
\end{aligned}\end{equation*}
Computing the variances using Eq.~\eqref{eq:4wm}, the noise figure
corresponding to four-wave mixing followed by optical losses is
\begin{equation}
    \mbox{NF} = 1 + \frac{2(G-1)(G(\eta_a-\eta_b)^2-\eta_b^2)}{G \eta_a + (G-1)\eta_b }.
    \label{eq:noabsorp}
\end{equation}
This expression highlights the importance of balanced beam
detection, as unbalanced losses ($\eta_a\neq\eta_b$) result in
detection of amplified noise instead of squeezing.

\section{Optical losses during squeezing: Interleaved gain/loss model}

The four-wave mixing process consists of Raman transitions between
the hyperfine ground states (Fig.~\ref{fig:fwm}A), which are most
efficient when the intermediate virtual level is tuned close to
resonance. However, this also increases direct absorption from the
Doppler broadened transition, increasing losses and reducing
correlations. To analyze this trade-off, we develop a model for
the effect of competing mixing and absorption on relative
intensity squeezing.

\begin{figure}[!hbt]\centering
    \includegraphics[width=0.9\textwidth]{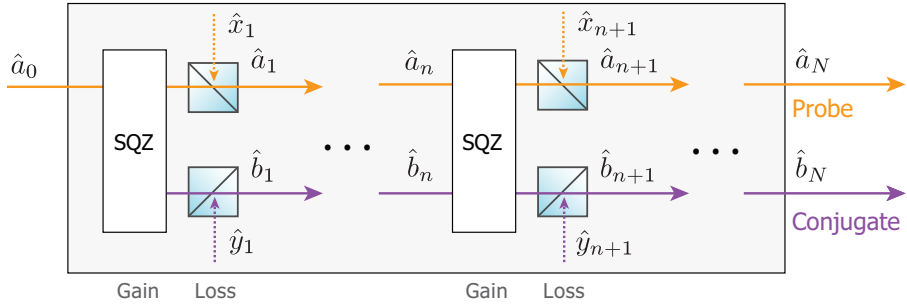}
\caption{Competing gain and loss processes modelled by interleaved
stages of squeezing (SQZ) and loss.}
    \label{fig:glmodel}
\end{figure}

Following the approach of the numerical model presented in
Ref.~\cite{lett_low}, the beam trajectories through the medium are
divided into $N$ discrete interleaved stages of gain and loss
(Fig. \ref{fig:glmodel}). Distributed models of this type were
first proposed by Loudon~\cite{loudon85}, and applied by Caves and
Crouch~\cite{caves87} to model distributed squeezing losses in a
single-mode parametric amplifier. We present a fully analytical
model of the relative squeezing produced by this system in the
continuum limit, including losses on both beams.

Each stage comprises ideal squeezing (by parameter $s$) followed
by loss (represented by transmission coefficients $t_a$ on the
probe and $t_b$ on the conjugate). Consecutive stages of the model
are related by combining Eq.~\eqref{eq:4wm} with
Eq.~\eqref{eq:loss} to give
\begin{equation}\begin{aligned}
    \op{a}_{n+1} &= t_a (\cosh s \;\op{a}_n + \sinh s \;\opd{b}_n) + \sqrt{1-t_a^2}\;\op{x}_{n+1}, \\
    \opd{b}_{n+1} &= t_b (\sinh s \;\op{a}_n + \cosh s \;\opd{b}_n) + \sqrt{1-t_b^2}\;\opd{y}_{n+1}.
    \label{eq:gltransf}
\end{aligned}\end{equation}
The overall model is parameterized in terms of the overall
squeezing parameter $S$ in the absence of losses, and transmissions
$T_a$ and $T_b$ in the absence of squeezing. These are related to
the incremental coefficients above by $s=S/N$, $t_a = T_a^{1/2N}$ and $t_b =
T_b^{1/2N}$.

This transformation can be written in matrix form and applied
recursively to express the output beam operators $\op{a}_N$ and
$\opd{b}_N$ as a sum of the incident beam operators ($\op{a}_0$ and
$\opd{b}_0$) and the injected vacuum operators ($\op{x}_i$ and
$\opd{y}_i$) as
\begin{equation}\begin{aligned}[b]
    \twovec{\op{a}_N}{\opd{b}_N} &= \mat{A} \twovec{\op{a}_{N-1}}{\opd{b}_{N-1}} + \twovec{\sqrt{1-t_a^2} \;\op{x}_N}{\sqrt{1-t_b^2}\;\opd{y}_N} \\
    &= \mat{A}^N \twovec{\op{a}_0}{\opd{b}_0} + \sum_{i=1}^N \mat{A}^{N-i} \twovec{ \sqrt{1-t_a^2}\; \op{x}_i}{\sqrt{1-t_b^2}\; \opd{y}_i } ,
\end{aligned}\label{eq:glmatrix}
\end{equation}
where
\begin{equation}
    \mat{A}=\begin{pmatrix} t_a \cosh s & t_a \sinh s \\ t_b \sinh s & t_b \cosh s \end{pmatrix} .
    \label{eq:matA}
\end{equation}
For algebraic simplicity, we introduce a polymorphic operator
$\op{z}_i$ consisting of the probe annihilation and conjugate
creation operators $\op{z}_1 = \op{a}_0$ and $\op{z}_2 =
\opd{b}_0$, followed by the injected vacuum operators
$\op{z}_{2i+1} = \op{x}_i$ and $\op{z}_{2i+2} = \opd{y}_i$ for
$1\le i \le N$. Expanding Eq.~\eqref{eq:glmatrix} in terms of this
operator and a set of coefficients $\alpha_i$ and $\beta_i$ gives
\begin{equation}\begin{aligned}
    \op{a}_N &= \alpha_1 \op{a}_0 + \alpha_2 \opd{b}_0 + \alpha_3 \op{x}_1 + \alpha_4 \opd{y}_1 + \cdots + \alpha_{2N+1} \op{x}_N + \alpha_{2N+2} \opd{y}_N &= \sum\nolimits_i \alpha_i \op{z}_i ,\\
    \opd{b}_N &= \beta_1 \op{a}_0 + \beta_2 \opd{b}_0 + \beta_3 \op{x}_1 + \beta_4 \opd{y}_1 + \cdots + \beta_{2N+1} \op{x}_N + \beta_{2N+2} \opd{y}_N &= \sum\nolimits_i \beta_i \op{z}_i .
\end{aligned}\label{eq:zexpand}\end{equation}
Using standard statistical identities, the relative intensity
operator and its variance are
\begin{equation*}\begin{aligned}
    \op{N}_a-\op{N}_b &= \opd{a}_N\op{a}_N - \opd{b}_N\op{b}_N = \sum\nolimits_{i,j} (\alpha_i\alpha_j - \beta_i\beta_j)\opd{z}_i\op{z}_j + 1
        \quad\text{and}\\
    \var{\op{N}_a-\op{N}_b } &= \sum\nolimits_{i,j,k,l} (\alpha_i\alpha_j - \beta_i\beta_j)(\alpha_k\alpha_l - \beta_k\beta_l)
        \; \cov{\opd{z}_i\op{z}_j}{\opd{z}_k\op{z}_l} .
\end{aligned}\end{equation*}
In the bright probe beam approximation ($\op{N}_0\gg1$), this
variance simplifies to
\begin{equation*}
    \var{\op{N}_a-\op{N}_b } = (\alpha_1^2 - \beta_1^2)^2\var{\op{N}_0} + \sum\nolimits_{i>1}(\alpha_1\alpha_i - \beta_1\beta_i)^2\expect{\op{N}_0} .
\end{equation*}
The shot-noise limit is $\expect{\op{N}_a+\op{N}_b} = (\alpha_1^2
+ \beta_1^2)\expect{\op{N}_0}$, so the degree of squeezing for an
initially shot-noise limited probe with
$\var{\op{N}_0}=\expect{\op{N}_0}$ is
\begin{equation}
    \text{NF} \equiv \frac{\var{\op{N}_a-\op{N}_b}}{\expect{\op{N}_a+\op{N}_b}} = \frac{\sum_{i=1}^N (\alpha_1\alpha_i - \beta_1\beta_i)^2}{\alpha_1^2 + \beta_1^2} .
    \label{eq:glnf}
\end{equation}
It remains to express the $\alpha_i$, $\beta_i$ coefficients in
terms of the model parameters $T_a$, $T_b$ and $S$, and hence
obtain an ab-initio expression for the degree of squeezing.

Equating the coefficients of Eq.~\eqref{eq:glmatrix} and
Eq.~\eqref{eq:zexpand} leads to
\begin{equation*}
    \begin{pmatrix} \alpha_1 & \alpha_2 \\ \beta_1 & \beta_2 \end{pmatrix} = \mat{A}^N
    \qquad\text{and}\qquad
    \begin{pmatrix}\alpha_{2i+1} & \alpha_{2i+2} \\ \beta_{2i+1} & \beta_{2i+2} \end{pmatrix} = \mat{A}^{N-i} \begin{pmatrix} \sqrt{1-t_a^2} & 0 \\ 0 & \sqrt{1-t_b^2} \end{pmatrix}.
    \label{eq:coeffeq}
\end{equation*}
Hence each of the $N$ vacuum modes $\op{x}_i$ contribute a term to
the variance in Eq.~\eqref{eq:glnf}:
\begin{equation}\begin{aligned}[b]
(\alpha_1 \alpha_{2i+1} - \beta_1 \beta_{2i+1})^2
    &= \tworvec{\alpha_1}{-\beta_1}\twovec{\alpha_{2i+1}}{\beta_{2i+1}}\tworvec{\alpha_{2i+1}}{\beta_{2i+1}}\twovec{\alpha_1}{-\beta_1} \\
    &= \tworvec{\alpha_1}{-\beta_1}\mat{A}^{N-i} \twovec{\sqrt{1-t_a^2}}0 \tworvec{\sqrt{1-t_a^2}}0\left(\mat{A}^T\right)^{N-i}\twovec{\alpha_1}{-\beta_1} \\
    &= \tworvec{\alpha_1}{-\beta_1}\mat{A}^{N-i} \begin{pmatrix} 1-t_a^2 & 0 \\ 0 & 0 \end{pmatrix} \mat{A}^{N-i}\twovec{\alpha_1}{-\beta_1} .
\end{aligned}\label{eq:vaccontrib}\end{equation}
Each operator $\opd{y}_i$ contributes a term analogous to
Eq.~\eqref{eq:vaccontrib}, but whose diagonal matrix is
\mbox{$\lceil 0 \quad 1-t_b^2 \rfloor$}. Temporarily neglecting
the $i=1,2$ contributions to the variance in Eq.~\eqref{eq:glnf},
and summing over the vacuum contributions gives
\begin{equation}
    \sum_{i>2}(\alpha_1 \alpha_i - \beta_1 \beta_i)^2 = \tworvec{\alpha_1}{-\beta_1}\left\lbrace \sum_{i=1}^N \mat{A}^{N-i} \begin{pmatrix} 1-t_a^2 & 0 \\ 0 & 1-t_b^2 \end{pmatrix} \mat{A}^{N-i}\right\rbrace\twovec{\alpha_1}{-\beta_1} .
    \label{eq:vacsum}
\end{equation}
The continuum behaviour is recovered in the limit
$N\rightarrow\infty$. To obtain a closed form expression for the
sum, the infinitesimal parameters are expanded as a power series
in $1/N$. Expanding the elements of $\mat{A}$ in
Eq.~\eqref{eq:matA} gives
\begin{equation}
    \mat{A} = 1 + \tfrac1N \mat{A}_0 + \ord1{N^2}   \quad \text{ where } \quad \mat{A}_0 =  \begin{pmatrix} \tfrac12 \log T_a & S \\ S & \tfrac12 \log T_b \end{pmatrix} .
    \label{eq:aexpand}
\end{equation}
Similarly taking $t_a^2=\exp(\frac1N\log T_a)\simeq1+\frac1N\log
T_a$, the sum in braces in Eq.~\eqref{eq:vacsum} is
\begin{equation*}
    \mat{X} = \sum_{i=1}^N \mat{A}^{N-i} \left(\tfrac1N{\mat{T}}\right) \mat{A}^{N-i} = \frac1N\sum_{i=0}^{N-1} \mat{A}^i \;\mat{T}\; \mat{A}^i     \qquad\text{ with }\qquad \mat{T} = \begin{pmatrix} -\log T_a & 0 \\ 0 & -\log T_b \end{pmatrix} .
\end{equation*}
It can be easily verified that this sum obeys the geometric series
relation
\[ \mat{A}\,\mat{X}\,\mat{A}-\mat{X} = \tfrac1N(\mat{A}^N\;\mat{T}\;\mat{A}^N-\mat{T}). \]
Expanding to order $1/N$ using Eq.~\eqref{eq:aexpand} gives
\begin{equation*}
    \mat{A}\,\mat{X}\,\mat{A}-\mat{X} = \tfrac1N \left\lbrace \mat{A}_0\;\mat{X} + \mat{X}\;\mat{A}_0 \right\rbrace \quad\Rightarrow\quad \mat{A}_0\,\mat{X} + \mat{X}\,\mat{A}_0 = \mat{A}^N\;\mat{T}\;\mat{A}^N - \mat{T} .
\end{equation*}
Taking the limit $N\rightarrow\infty$, the neglected $O(1/N^2)$
terms vanish and $\mat{A}^N\rightarrow\exp(\mat{A}_0)$. Hence the
sum $\mat{X}$ converges and obeys
\begin{equation}
    \mat{A}_0\,\mat{X} + \mat{X}\,\mat{A}_0 = \exp(\mat{A}_0) \;\mat{T}\; \exp(\mat{A}_0) - \mat{T}.
\end{equation}
This is a system of four linear equations for the elements of
$\mat{X}$ in terms of the model parameters $T_a$, $T_b$ and $S$,
and can be solved algebraically.

The sum $\mat{X}$ contains all terms in the variance of
Eq.~\eqref{eq:glnf} except $i=1,2$ which correspond to the probe
and conjugate coefficients. The probe contribution is
\begin{equation*}
    (\alpha_1^2 - \beta_1^2)^2 = \left\lbrace\tworvec{\alpha_1}{-\beta_1}\twovec{\alpha_1}{\beta_1}\right\rbrace^2 = \tworvec{\alpha_1}{-\beta_1} e^{\mat{A}_0} \begin{pmatrix}1&0\\0&0\end{pmatrix} e^{\mat{A}_0} \twovec{\alpha_1}{-\beta_1} ,
\end{equation*}
while the conjugate contribution
$(\alpha_1\alpha_2-\beta_1\beta_2)^2$ has diagonal \mbox{$\lceil0
\quad 1\rfloor$}. Computing the full variance sum in
Eq.~\eqref{eq:glnf} yields
\begin{equation*}
    \var{\op{N}_a-\op{N}_b} =  \tworvec{\alpha_1}{-\beta_1} \Bigl\lbrace e^{2\mat{A}_0} + \mat{X} \Bigr\rbrace \twovec{\alpha_1}{-\beta_1}
        \quad\text{with}\quad \twovec{\alpha_1}{-\beta_1} = \begin{pmatrix}1&0\\0&-1\end{pmatrix} e^{\mat{A}_0} \twovec10 .
\end{equation*}
We introduce one final stage of loss to model optical losses after
mixing (as in \S4), scaling each coefficient by the relevant
transmission factor ($\sqrt{\eta_a}$ or $\sqrt{\eta_b}$) and
introducing the extra vacuum contributions
$\eta_a(1-\eta_a)\alpha_1^2 + \eta_b(1-\eta_b)\beta_1^2$. The net
variance in Eq.~\eqref{eq:glnf} is therefore
\begin{equation*}
    \var{\op{N}_a-\op{N}_b} =  \tworvec{\alpha_1}{-\beta_1} \Bigl\lbrace \mat{P}(e^{2\mat{A}_0}+\mat{X})\mat{P} + (1 - \mat{P})\mat{P} \Bigr\rbrace \twovec{\alpha_1}{-\beta_1}\expect{\op{N}_0}
        \;\;\text{where}\;\; \mat{P}=\begin{pmatrix} \eta_a & 0 \\ 0 & \eta_b \end{pmatrix}.
\end{equation*}
The measured beam powers relative to the incident probe power (the
``effective gains'') are
\begin{equation}
    G_a \equiv \frac{\expect{\op{N}_a}}{\expect{\op{N}_0}} = \eta_a \alpha_1^2      \quad\text{ and }\quad      G_b \equiv \frac{\expect{\op{N}_b}}{\expect{\op{N}_0}} = \eta_b \alpha_2^2
        \quad\text{ with }\quad \twovec{\alpha_1}{\alpha_2} =  e^{\mat{A}_0}\twovec10,
    \label{eq:egains}
\end{equation}
and the relevant shot-noise limit is $
\var{\op{N}_a-\op{N}_b}_{\text{SNL}} =
\expect{N_a}+\expect{N_b}=(\eta_a \alpha_1^2 + \eta_b
\beta_1^2)\expect{\op{N}_0} $.

Evaluating all the contributions to Eq.~\eqref{eq:glnf}, we obtain
an analytic expression for the degree of relative intensity
squeezing produced by this system. This expression is algebraic in
$T_a$, $T_b$ and $S$, but runs to a dozen typeset lines. However,
special cases are readily derived and provide physical insight not
readily accessible from numerical models.

In the experimentally studied case (Refs. [1--7]), detection
efficiencies are carefully balanced ($\eta_a=\eta_b\equiv\eta$)
and the far-detuned conjugate experiences negligible absorption
($T_b=1$). The corresponding degree of squeezing is
\begin{equation}
    \text{NF} = 1 - \eta \frac{2S \sinh^2\!\xi}{\xi \cosh(2\xi+\chi)} + \eta\sqrt{T_a}\frac{S \: \log^2\!T_a  \sinh^4\! \xi}{2 \xi^3 \cosh(2\xi+\chi)} ,
    \label{eq:tb1}
\end{equation}
with parameters $\xi=\frac14\sqrt{16S^2+(\log T_a-\log T_b)^2}$
and $\tanh\chi = (\log T_a-\log T_b)/4\xi$. The three terms
describe the shot-noise limit, correlations from four-wave mixing,
and injected vacuum noise.

\begin{figure}[!bt]\centering
    \includegraphics[scale=0.7]{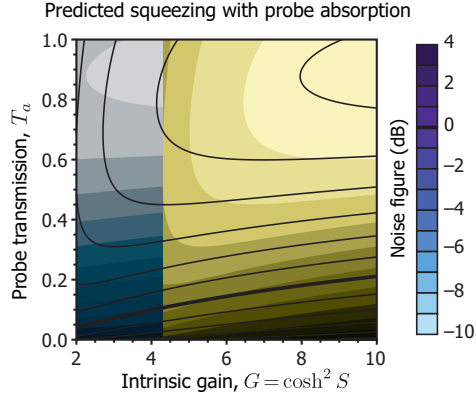}
    \caption{Predicted squeezing for four-wave mixing with negligible conjugate absorption and net detection efficiency $\eta=85\%$.}
    \label{fig:compareloss}
\end{figure}

Figure~\ref{fig:compareloss} shows the noise figure as a function
of the probe transmission $T_a$ and intrinsic mixing gain
$G=\cosh^2S$ (in excellent agreement with the numerical
model of Ref.~\cite{lett_low}). Note the counter-intuitive result
that the strongest squeezing is obtained with imperfect
transmission ($T_a<1$). This is because the shot noise carried by
the incident probe beam is also amplified by the mixing process,
and a small ``optimal'' level of probe loss decreases this
contribution to the measured noise power before injected vacuum
noise dominates. This optimal level is easily obtained by
minimizing Eq.~\eqref{eq:tb1}.

\begin{figure}[!bt]\centering
    \includegraphics[width=\textwidth]{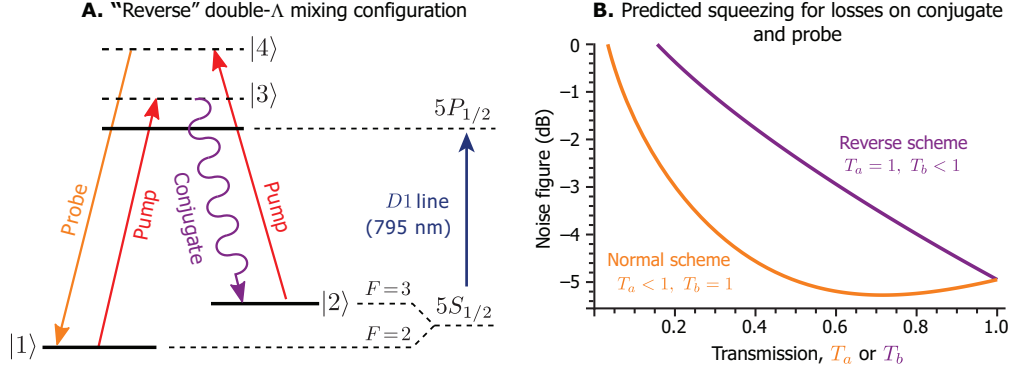}
    \caption{(A) Four-wave mixing in the ``reverse'' configuration. (B) Predicted squeezing for four-wave mixing in the normal/reverse configurations with intrinsic gain $G=3$.}
    \label{fig:reverse}
\end{figure}

The model can also be applied to other four-wave mixing systems.
For example, interchanging the probe and conjugate wavelengths
produces the ``reverse'' configuration (Fig.~\ref{fig:reverse}A).
This system is interesting as the weakly-coupled conjugate
transition is brought closer to resonance, producing much higher
intrinsic mixing gain for the same beam powers. In this case, the
probe experiences negligible absorption ($T_a=1$) compared to the
conjugate ($T_b<1$), and the predicted squeezing is
\begin{equation}
    \text{NF} = 1 - \eta \frac{2S\cosh^2(\xi+\chi)}{\xi \cosh(2\xi+\chi)} + \eta\sqrt{T_b} \frac{S(4S - \log T_b \sinh(2\xi+\chi))^2}{8 \xi^3 \cosh(2\xi +\chi)} .
    \label{eq:ta1}
\end{equation}
The vacuum noise term in Eq.~\eqref{eq:ta1} is considerably larger
than in Eq.~\eqref{eq:tb1}, resulting in several decibels
difference for moderate levels of absorption (Fig.
\ref{fig:reverse}B). Unlike the probe losses discussed above,
losses on the conjugate only destroy correlations and introduce
noise, so squeezing by four-wave mixing in the reverse
configuration is always less effective for the same level of
intrinsic mixing gain.

Finally, it is worth noting for consistency that in the limit of
both $T_a$, $T_b\rightarrow1$, the post-mixing optical-loss result
of Eq.~\eqref{eq:noabsorp} is obtained.

\section{Experimental diagnostic}\label{sec:expr}
The analytic model derived above provides a simple yet powerful
tool for optimizing the degree of squeezing obtained
experimentally. It not only predicts which parameters will provide
optimal results, but can also be used as a diagnostic tool to
determine which factors are limiting the experimentally measured
degree of squeezing.

To demonstrate this, we constructed a four-wave mixing apparatus
as described in Ref.~\cite{lett07}. A $400\,\text{mW}$ pump beam
intersects a $100\,\mu\text{W}$ probe beam with $1/e^2$ beam
waists of $630\,\mu$m and $375\,\mu$m respectively at an angle of
$0.3^\circ$ within a pure $^{85}$Rb vapor cell of internal length
$7\,\text{mm}$ heated to $130^\circ$C (Fig. \ref{fig:fwm}B). The
probe was generated by an AOM with fixed detuning $3040\,$MHz
below the pump, which was scanned across the Doppler broadened
$D_1$ resonance at $795\,$nm. The relative intensity between probe
and conjugate was measured with a balanced photodetector (Thorlabs
PDB150A), refitted with high-efficiency photodiodes (Hamamatsu
S3883, net efficiency $95\%$). The overall detection efficiency of
the system was $\eta=85\pm1\%$. The relative intensity noise was
measured with a Rhode \& Schwarz FSP7 spectrum analyzer at an
analysis frequency of $1\,$MHz with $30\,$kHz resolution
bandwidth.

\begin{figure}[!bt]\centering
    \includegraphics[width=\textwidth]{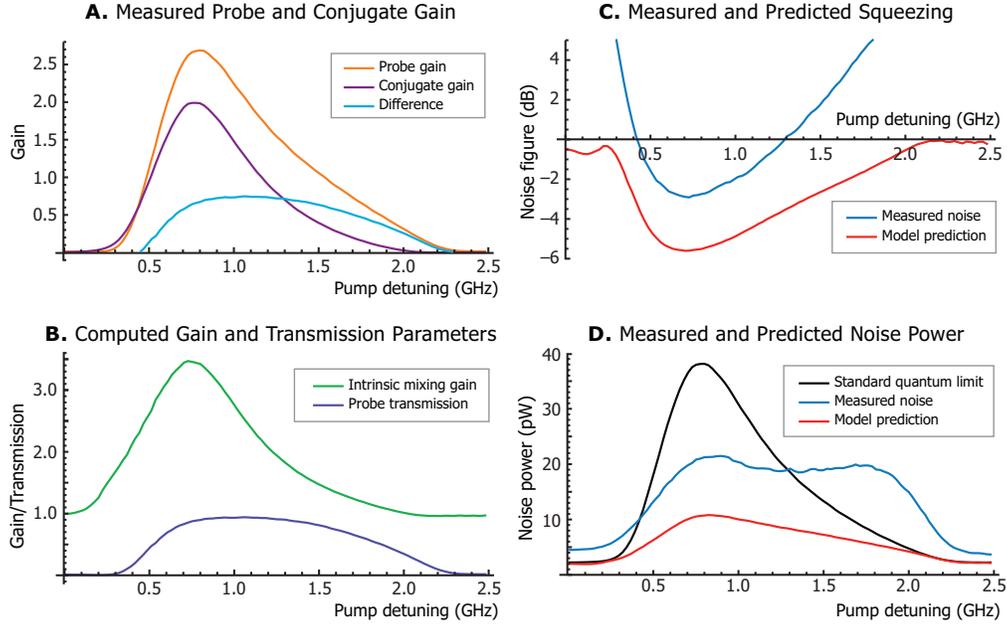}
    \caption{(A) Measured probe and conjugate gain across four-wave mixing resonance and (B) inferred model parameters. (C) Experimentally measured squeezing and (D) associated noise powers compared to model predictions for these parameters. Detunings are for pump beam and are measured above the centre of the $^{85}$Rb $5S_{1/2} (F=2)\rightarrow 5P_{1/2}$ transition.}
    \label{fig:exper}
\end{figure}

The effective probe and conjugate gains ($G_a$ and $G_b$) were
measured as a function of pump beam detuning (Fig.
\ref{fig:exper}A) and used to simultaneously solve
\mbox{Eq.~\eqref{eq:egains}} for the intrinsic gain $G$ and probe
transmission $T_a$ (Fig. \ref{fig:exper}B) via the coefficients
$\alpha_1$ and $\alpha_2$. Note that the gain resonance extends
well into the Doppler-broadened absorption resonance for detunings
below $600\,$MHz from the line-centre, demonstrating strong
competition between the processes.

The model predicts significantly stronger squeezing should be
possible for these parameters than measured experimentally (Fig.
\ref{fig:exper}C). As our incident probe beam was measured to be
shot-noise limited, this demonstrates that our noise measurement
was limited by technical factors and not by insufficient intrinsic
gain or excessive absorption.

One cause for our discrepancy is the technical difficulty in
eliminating the bright pump beam after the vapor cell. Detection
of the pump introduces uncorrelated fluctuations and increases the
measured noise level~\cite{lett_low}. This is evident around zero
detuning, where the probe is fully absorbed (Fig.
\ref{fig:exper}A) but the measured noise is far above the standard
quantum limit (Fig. \ref{fig:exper}D). Upon blocking the pump
beam, the relative intensity noise was measured at the shot-noise
limit. Subtracting this background level implies that $-4.1\,$dB
of squeezing would be obtained by eliminating cross-beam
detection.

The model assumes that gain occurs uniformly throughout the vapor
cell, which requires that the beams be overlapped over the entire
region. If the beams are not properly overlapped, the mixing
strength decreases and gain becomes spatially varying. As losses
are unchanged, the relative intensity noise increases as a result.
This is likely the cause of the remaining discrepancy, and
improved squeezing could be achieved by manipulating beam
alignment and waists to ensure proper overlap throughout the cell.

\section{Conclusions}
We have presented a method for analytically calculating the degree
of squeezing produced by a four-wave mixing system in the presence
of absorption. Our model included the contributions from
arbitrarily many injected vacuum modes that were subsequently
squeezed by the system, producing an ab-initio quantum mechanical
description of the introduced losses. Our general result is not
reliant on implementation details and can be applied to analyze
any four-wave mixing scheme, while the matrix methods techniques
we developed can be applied to model other systems both within
quantum optics and more generally.

We considered two special cases, corresponding to the
experimentally studied system (introduced in Ref.~\cite{lett07})
and the ``reverse'' configuration, with probe and conjugate
wavelengths interchanged. We presented closed-form expressions for
the relative intensity noise in these cases, and demonstrated that
a small level of probe loss was desirable to suppress
amplification of the initial shot-noise. The reverse configuration
was shown to produce the same squeezing at equal gain for ideal
transmission, but was significantly more sensitive to losses.

The model was applied to analyze experimentally measured squeezing
and determine the intrinsic mixing gain and transmission factors
of the four-wave mixing resonance. Comparing the expected
squeezing to measured results provided insight into the factors
limiting our measurement and hence where to direct effort in
optimizing the many free parameters of the system.

While it should be noted that this model considers the propagation
of a Gaussian beam mode only, an arbitrary beam can be analyzed as
a product state of orthogonal spatial modes, with each mode
independently squeezed~\cite{lett_qimaging}. The model established
in this paper may therefore be applied to each pair of spatial
modes and the resulting noise powers summed to obtain the overall
relative intensity squeezing for a multi-mode beam. Such
multi-mode squeezing has been experimentally
demonstrated~\cite{lett_science}, with promising applications in
quantum imaging.

\end{document}